\documentclass[12p]{article}
\usepackage{latexsym}
\usepackage{color}
\usepackage{amsmath}
\usepackage{epsfig}
\usepackage{hyperref}
 \usepackage{dcolumn}
\newcommand{\ortala}[1]{\begin{center}#1\end{center}}

\newcommand{\sandd}[1]{\left\langle #1\right\rangle}

\newcommand{\intego}[3]{{{\underset{#1
}{\overset{#2}{\displaystyle\oint}}}#3}}
\newcommand{\summ}[3]{{{\underset{#1 }{\overset{#2}{\displaystyle\sum}}}#3}}

\newcommand{\re}[1]{(\ref{#1})}

\newcommand{\eq}[2]{\begin{equation}\label{#1}  #2\end{equation}}
\newcommand{\tur}[2]{\frac{d#1}{d#2}}

\newcommand{\paran}[1]{\left(#1\right)}

\newcommand{\sch}[1]{Schrodinger}

\linethickness{0.6pt}

\begin{document}

\ortala{\textbf{ Nonequilibrium phase transitions in isotropic Ashkin-Teller 
Model}}

\ortala{\textbf{\"Umit Ak\i nc\i \footnote{umit.akinci@deu.edu.tr}}}

\ortala{\textit{Department of Physics, Dokuz Eyl\"ul University,
TR-35160 Izmir, Turkey}}

\section{Abstract}

Dynamic behavior of an isotropic Ashkin-Teller model  in the presence of a 
periodically
oscillating magnetic field has been analyzed by means of the mean field 
approximation.
The dynamic equation of motion has been constructed with the
help of a Glauber type stochastic process and solved for a square  lattice.

After defining the possible dynamical phases of the system,  phase diagrams have been
given and the behavior of the hysteresis
loops has been investigated in detail.   The hysteresis loop for specific order 
parameter of isotropic
Ashkin-Teller model has been defined and characteristics of this loop in 
different dynamical phases have been given.

Keywords: \textbf{Dynamic isotropic Ashkin-Teller model;
hysteresis loops; hysteresis loop area}

\section{Introduction}\label{introduction}

Ashkin-Teller Model (ATM) has been introduced for description of the cooperative 
phenomena of quaternary
alloys \cite{ref1}. It has four states per site and may be useful to describe 
magnetic systems with two easy axes.
The ATM is a ‘staggered’ version of the eight-vertex model \cite{ref2}. In two 
dimension, the ATM can be mapped onto 
a staggered eight-vertex model at the critical point. The non-universal
critical behavior along a self-dual line, where the exponents vary continuously 
\cite{ref3},
is one of the interesting critical property of the model. On the other hand it 
has been shown that, 
three dimensional model has  much richer phase diagrams than the ATM in two 
dimension \cite{ref4}. 
There appear some first-order phase transitions and continuous phase 
transitions,
even an XY -like transition and a Heisenberg-like multicritical point.

It has been shown that ATM could be described in Hamiltonian form appropriate 
for spin systems \cite{ref5}.  In this form, the model can be viewed as two 
coupled Ising models which is named as 2-color ATM. If two of these Ising models 
are identical then the model named as isotropic Ashkin-Teller model (IATM), 
otherwise the model is anisotropic Ashkin-Teller model (AATM). In a similar 
manner,  ATM that formed by $N$ coupled Isig model entitled as N-color ATM as 
introduced in \cite{ref6}.

One of the well known physical realizations for this model is the compound of 
Selenium adsorbed on a Ni
surface \cite{ref7}. ATM can be used to describe chemical interactions in 
metallic alloys \cite{ref8},
thermodynamic properties in superconducting cuprates (ATM represents the 
interactions between orbital current loops 
in $CuO_2$ -plaquettes) \cite{ref9} and elastic response of DNA molecule to
external force and torque \cite{ref10}. Besides, oxygen ordering in $YBa_2 Cu_3 
O_z$
may also be understood in analogy with the two-dimensional IATM 
\cite{ref11,ref12,ref13}. Also, ATM has many interesting applications in neural 
networks \cite{ref14} and cosmology \cite{ref15}. Besides, some mappings between 
the ATM and some other models are possible. This makes the ATM valuable in a 
theoretical manner. For instance, the random N-color quantum ATM can be 
described by an $O(N)$ Gross-Neveu model with random mass \cite{ref16}. 
Similarly the relation between the two-dimensional N-component Landau-Ginzburg 
Hamiltonian with cubic anisotropy and N-color ATM has been discussed in 
\cite{ref17,ref18}.

Critical properties of the ATM has been widely investigated in literature.  
IATM has been investigated by mean field renormalization group technique 
\cite{ref19,ref20}, Monte Carlo Simulation (MC) 
\cite{ref21,ref22,ref23,ref24,ref25,ref26,ref27}, effective field theory (EFT) 
\cite{ref28}, transfer-matrix finite-size-scaling method \cite{ref29}, high and 
low temperature series expansion and MC \cite{ref30}, MC and renormalization 
group technique \cite{ref31,ref32} and damage spreading simulation 
\cite{ref33,ref34,ref35}.

On the other hand, the AATM has been investigated within several techniques such 
as, mean field approximation (MFA) \cite{ref36}, MC \cite{ref37,ref38},  real-space 
renormalization-group approach \cite{ref39} and Monte Carlo renormalization 
group technique \cite{ref40}.

Theoretical works devoted to ATM is not limited to regular (translationally 
invariant lattices) lattices. For instance, the model has been solved on Bethe 
Lattice \cite{ref41}, diamond-like hierarchical lattice \cite{ref42,ref43} and 
Cayley tree \cite{ref44}. 

Also, some extensions and variants of the model have been studied. Extended ATM 
(ATM with reduced degeneracy) has been introduced \cite{ref46} and solved within 
the MFA \cite{ref47}, mean field renormalization group technique 
\cite{ref48} and MC \cite{ref49}. Some other extensions like ATM with 
Dzyaloshinskii –Moriya interaction \cite{ref50}, ATM with spin-1 
\cite{ref51,ref52}, as well as mixed spin ATM \cite{ref45} can be found in the 
literature. 

Likewise, N-color ATM introduced in \cite{ref6}  has been investigated within 
the transfer matrix analysis \cite{ref53}, renormalization group technique 
\cite{ref54}, Monte Carlo renormalization group technique \cite{ref55}, MFA
and MC \cite{ref56}. This model has also been solved exactly 
for large N in 2D \cite{ref57,ref58}.

Although, magnetic systems under a time dependent external magnetic field has 
been attracted much interest from both theoretical and experimental points of 
view; to the best of our knowledge,  ATM under the time dependent magnetic field 
has not yet been studied. It will be interesting to investigate the dynamic 
character of ATM, which has rich critical properties in the static case, and 
obtain the corresponding dynamic phases of the system.  

Dynamic phase transition (DPT) in magnetic systems comes from the competition 
between the relaxation time of the system and period of the driving periodic 
external magnetic field  \cite{ref59}. The time average of the magnetization
over a full period of the oscillating magnetic field can be used as
dynamic order parameter (DOP) of the system. On the other hand appeared 
hysteresis behavior
from the delay of the response of the system to the driving cyclic force 
includes important clues of the dynamic character of the system.

From the experimental point of view, 
DPTs and hysteresis behaviors can be observed
experimentally in different types of magnetic systems.
Experiments on ultrathin Co films \cite{ref60},  Fe/Au(001)
films \cite{ref61}, epitaxial Fe/GaAs(001) thin
films \cite{ref62},  fcc Co(001), and fcc NiFe/Cu/Co(001) layers \cite{ref63}
Fe/InAs(001) ultrathin films \cite{ref64} are among them.

On the theoretical side, there has been growing interest in the DPT which was 
first observed within the MFA \cite{ref65} for the s-$1/2$ Ising model. Since that time,  
DPT and hysteresis behaviors of the s-$1/2$ Ising model have been widely studied 
within the
several techniques such as MFA \cite{ref66}, Monte Carlo 
simulation (MC) \cite{ref67},
effective field theory (EFT) \cite{ref68}. 

The aim of this work
is to investigate the IATM under a magnetic field
oscillating in time within the MFA formulation 
and Glauber type of stochastic process \cite{ref71}. For this aim the paper is 
organized
as follows: In Sec. \ref{formulation} we briefly present the
model and  formulation. The results and discussions are
presented in Sec. \ref{results}, and finally Sec. \ref{conclusion} contains our 
conclusions.

\section{Model and Formulation}\label{formulation}

The Hamiltonian of the dynamical IATM is given by
\eq{denk1}{\mathcal{H}=-J_2\summ{<i,j>}{}{\paran{\sigma_i \sigma_j+s_i s_j}}
-J_4\summ{<i,j>}{}{\sigma_i \sigma_j s_i s_j}-
H(t)\summ{i}{}{\paran{\sigma_i+s_i}},}
where the first two summations are over the nearest neighbors of the lattice, 
while the
last one is over all the lattice sites. Here, $\sigma_i$ and  $s_i$ are the $z$ 
components of the spin variables at a site $i$, $J_2$ is the bilinear exchange 
interactions  of both of the Ising models.   $J_4$ is the four spin interaction, 
which couples two Ising models to produce the IATM. Dynamical character of the 
model comes from the time dependent external longitudinal magnetic field $H(t)$ 
and it is given by
\eq{denk2}{H(t)=H_0\sin{\paran{\omega t}},}
where $H_0$ is the amplitude and $\omega$ is the angular frequency of
the periodic magnetic field and $t$ stands for the time. 

In the static case (i.e. $H_0=0$) the model turns to the IATM, which is 
equivalent to the four component Potts model for $J_2=J_4$ \cite{ref72}. When 
$J_4=0$ the model reduces to two independent Ising models. 

We use a Glauber-type stochastic process \cite{ref71} to
investigate dynamic properties of the considered system. In general, in the
Glauber type of stochastic process (as done in Ref. \cite{ref65} for the 
Ising model) the thermal average (denoted with $\sandd{}$) of a spin variable 
$\tau_i$, which can take values $\pm 1$  
can be given as
\eq{denk3}{
\theta \tur{\sandd{\tau_i}}{t}=-\sandd{\tau_i}+\sandd{\frac{Tr_i \tau_i 
\exp{\paran{-\beta \mathcal{H}_i}}}{Tr_i \exp{\paran{-\beta \mathcal{H}_i}}}}
}
in this type of process. Here, $\theta$ is the  transition rate per
unit time, $\beta=1/(k_BT)$, $k_B$ and $T$ denote the Boltzmann constant and
temperature, respectively. $Tr_i$ stands for the trace operation over the site 
$i$. Also $\mathcal{H}_i$ denotes the part of the Hamiltonian of the system 
related to the site $i$, which is given by,

\eq{denk4}{
\mathcal{H}_i=-J_2\sigma_i\summ{j}{}{ \sigma_j}-J_2s_i\summ{j}{}{ s_j}
-J_4\sigma_i s_i\summ{j}{}{ \sigma_j  s_j}-
H(t)\paran{\sigma_i+s_i}, 
} where all summations are carried over the nearest neighbor sites of the site 
$i$.

In order to handle these spin-spin interactions, usual approximation can be 
adopted, that is; all spin-spin interactions of these variables are represented by 
local fields as
\eq{denk5}{
h_1=J_2\summ{j=1}{z}{\sigma_j},\quad h_2=J_2\summ{j=1}{z}{s_j},\quad 
h_4=J_4\summ{j=1}{z}{\sigma_j s_j},
} where $z$ is the coordination number (i.e. number of nearest neighbor sites of 
any site $i$) of the lattice. Then one spin cluster Hamiltoian gets the form
\eq{denk6}{
\mathcal{H}_i=-\sigma_i h_1-s_i h_2
-\sigma_i s_i h_4-\paran{\sigma_i +s_i}H(t)
}

By writing Eq. \re{denk6} in Eq. \re{denk3} and performing the $Tr_i$ operations 
we can get equations for $\tau_i=\sigma_i,s_i$ and $\sigma_i s_i$ as
$$
\theta 
\tur{\sandd{\sigma_i}}{t}=-\sandd{\sigma_i}+\sandd{F\paran{x,y,z}|_{x=h_1+H(t),
y=h_2+H(t),z=h_4}}
$$
\eq{denk7}{
\theta 
\tur{\sandd{s_i}}{t}=-\sandd{s_i}+\sandd{F\paran{x,y,z}|_{x=h_2+H(t),y=h_1+H(t),
z=h_4}}
}
$$
\theta \tur{\sandd{\sigma_i s_i}}{t}=-\sandd{\sigma_i 
s_i}+\sandd{F\paran{x,y,z}|_{x=h_4,y=h_1+H(t),z=h_2+H(t)}}
$$ where
\eq{denk8}{
F\paran{x,y,z}=\frac{\tanh{(\beta x)}+\tanh{(\beta y)}\tanh{(\beta 
z)}}{1+\tanh{(\beta x)}\tanh{(\beta y)}\tanh{(\beta 	z)}}
}

The simplest way to one calculate Eq. \re{denk7} is adopt MFA. Within this approximation, all local fields defined in Eq. 
\re{denk5} written in terms of the thermal averages of the spin variables,
\eq{denk9}{
h_1=zJ_2m_\sigma,\quad h_2=zJ_2m_s,\quad h_4=zJ_4m_{\sigma s},
} where 
\eq{denk10}{
m_\sigma=\sandd{\sigma_j}, m_s=\sandd{s_j}, m_{\sigma s}=\sandd{\sigma_j s_j}.
} Here, the translationally invariance property of the lattice is adopted, i.e. 
all lattice sites are equivalent.   Then Eq. \re{denk7} gets the form
$$
\theta \tur{m_\sigma}{t}=-m_\sigma+F\paran{x,y,z}|_{x=h_1+H(t),y=h_2+H(t),z=h_4}
$$
\eq{denk11}{
\theta \tur{m_s}{t}=-m_s+F\paran{x,y,z}|_{x=h_2+H(t),y=h_1+H(t),z=h_4}
}
$$
\theta \tur{m_{\sigma s}}{t}=-m_{\sigma 
s}+F\paran{x,y,z}|_{x=h_4,y=h_1+H(t),z=h_2+H(t)}
$$ 
We can obtain explicit form of the 
dynamic equation of motion Eq. \re{denk11}, by writing local fields defined in 
Eq. \re{denk9} into the function Eq. \re{denk8} given with order in Eq. 
\re{denk11}. Eq. \re{denk11} is a  coupled first order differential equation 
system and can be solved with standart methods such as Runge-Kutta method 
\cite{ref73}. This iterative method starts by some initial values of the order 
parameters ($m_\nu(0)$, $\nu=\sigma,s,\sigma s$) and results in the desired solution after the convergency criteria 
$m_\nu\paran{t}=m_\nu\paran{t+2\pi/\omega}$ is satisfied.
By this way we can obtain DOP as
\eq{denk12}{Q_\nu=\frac{\omega}{2\pi}\intego{}{}{m_\nu\paran{t}dt}} where 
$m_\nu\paran{t}$
is a stable and periodic function. Dynamical critical points can be determined 
by obtaining the variation of the $Q_\nu$
with temperature for given set of Hamiltonian parameters.

We can construct the hysteresis loops which are nothing but the variation
of the $m_\nu(t)$ with $H(t)$ in one period of the periodic magnetic field.
Hereafter,  once the hysteresis loop is determined, some quantities about
it can be calculated. One of them is  dynamical hysteresis loop area (HLA) and 
can be calculated
via integration over a complete cycle of the magnetic field,
\eq{denk13}{A_\nu=\intego{}{}{}m_\nu\paran{t}dH} and corresponds to the energy
loss due to the hysteresis.

\section{Results and Discussion}\label{results}

In order to determine DPT and hysteresis characteristics of the system, let us 
scale all Hamiltonian parameters with  $J_2$, i.e. the unit of energy is $J_2$,
\eq{denk21}{K_4=\frac{J_4}{J_2}, \tau=\frac{k_BT}{J_2}, 
h_0=\frac{H_0}{J_2},h(t)=\frac{H(t)}{J_2}.
} 
Our investigation will be restricted to square  ($z=4$) lattice. 
We set $\theta=1$ throughout our numerical calculations. 
We note that, under the transformation $\sigma_i\rightarrow s_i$, Hamiltonian 
and the formulation used here do not change, then it will be enough to 
investigate only one of the $m_\sigma$ or $m_s$, due to the fact that 
$m_\sigma=m_s$.

Different phases of the static model is well known:
\begin{itemize}
\item Ferromagnetic (Baxter) phase: All magnetizations are different from zero.
\item Paramagnetic phase: All magnetizations are zero.
\item $\sandd{\sigma s}, \sandd{\sigma}, \sandd{ s}$ phases : All magnetizations 
are zero except the magnetization that have index which is the name of the 
phase. For instance in $\sandd{\sigma}$ phase only $m_\sigma$ is different from 
zero, while other two are equal to zero.  
\end{itemize}

We investigate corresponding dynamical phases in this work. Three different 
dynamical phases (corresponding to the phases of the static IATM) will manifest 
themselves. Also, as discussed in Ref. \cite{ref65} and successive works related to 
the DPT in Ising model, some regions may appear in the phase diagram, which is 
overlap region of the dynamically ordered and disordered phases. This overlap 
region mostly occur for relatively large values of $h_0$ for the Ising model. 
Let us enumerate these phases with prefix DP (which stands for dynamical phase).

\begin{itemize}
\item DP1: All DOPs are different from zero (corresponds to the Baxter phase in 
static the case )
\item DP2: All DOPs are equal to zero  (corresponds to the paramagnetic phase in 
static the case )
\item DP3: Only $m_{\sigma s}$ is different from zero while $m_\sigma=m_s=0$ 
(corresponds to the $\sandd{\sigma s}$ phase in static the case )
\item DP4: Overlap of the DP1 and DP2 (or DP3) phases. 
 \end{itemize}

\begin{figure}[h]\begin{center}
\epsfig{file=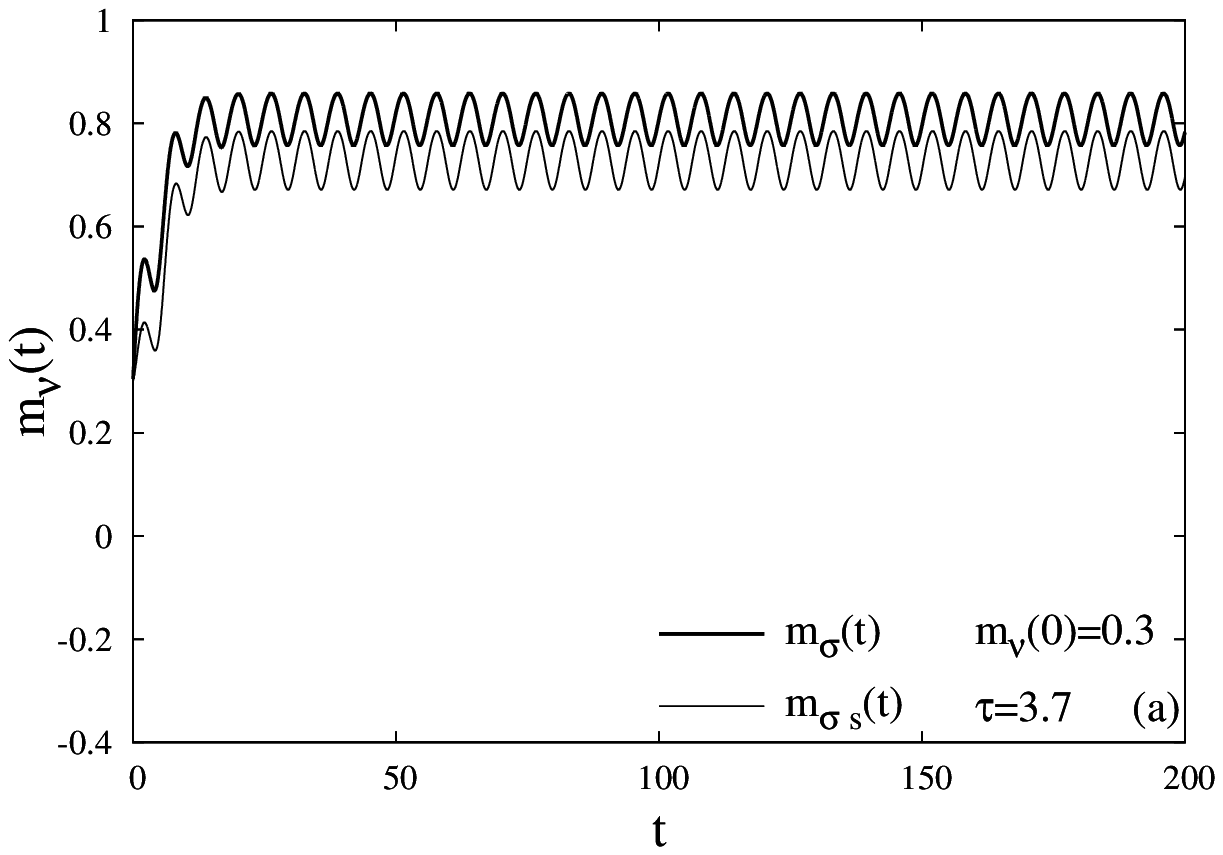, width=7.2cm}
\epsfig{file=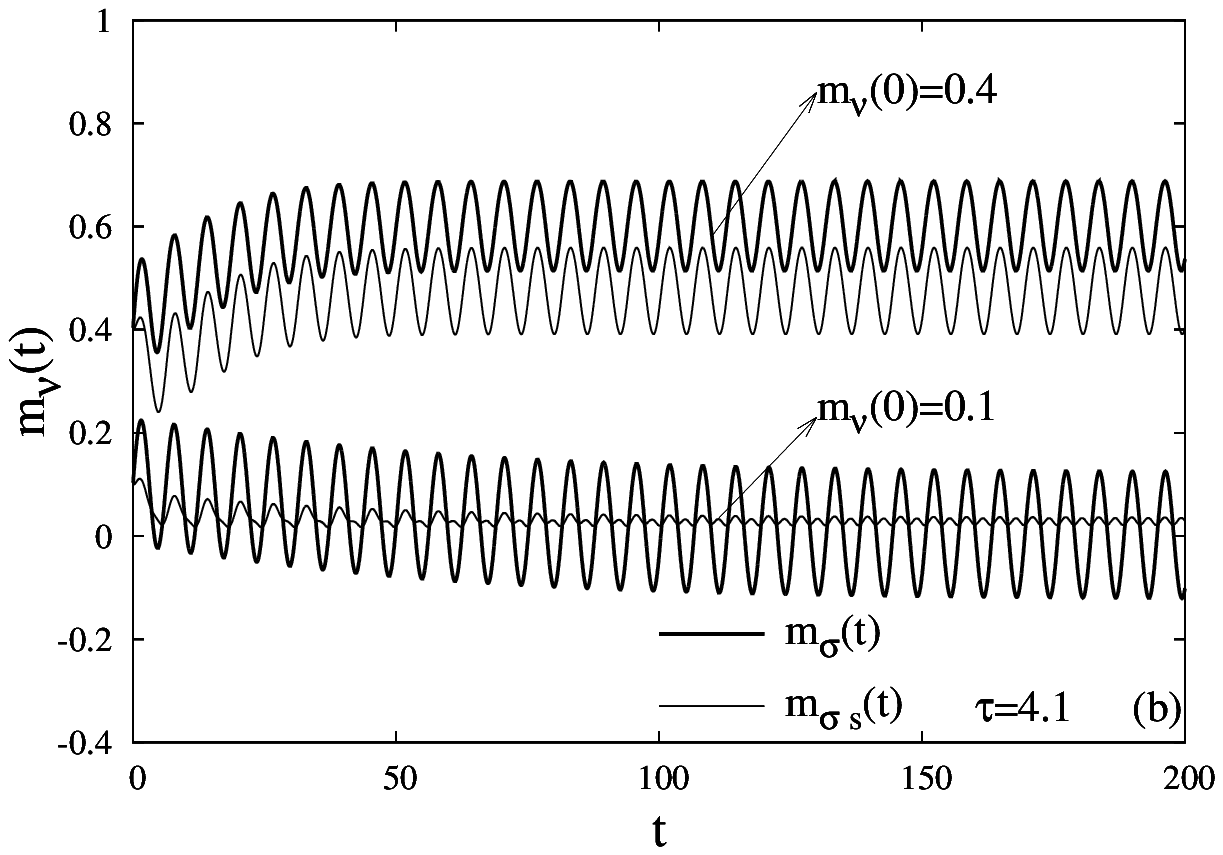, width=7.2cm}
\epsfig{file=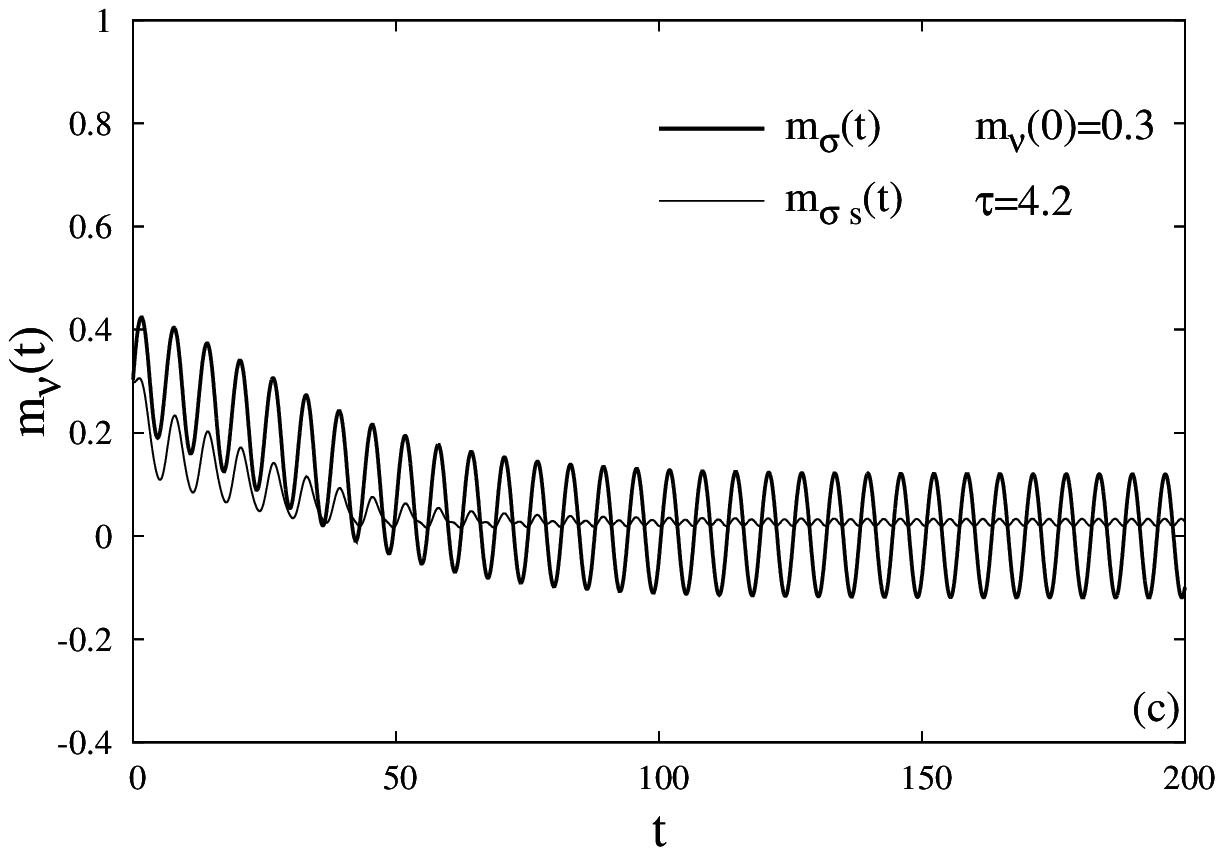, width=7.2cm}
\epsfig{file=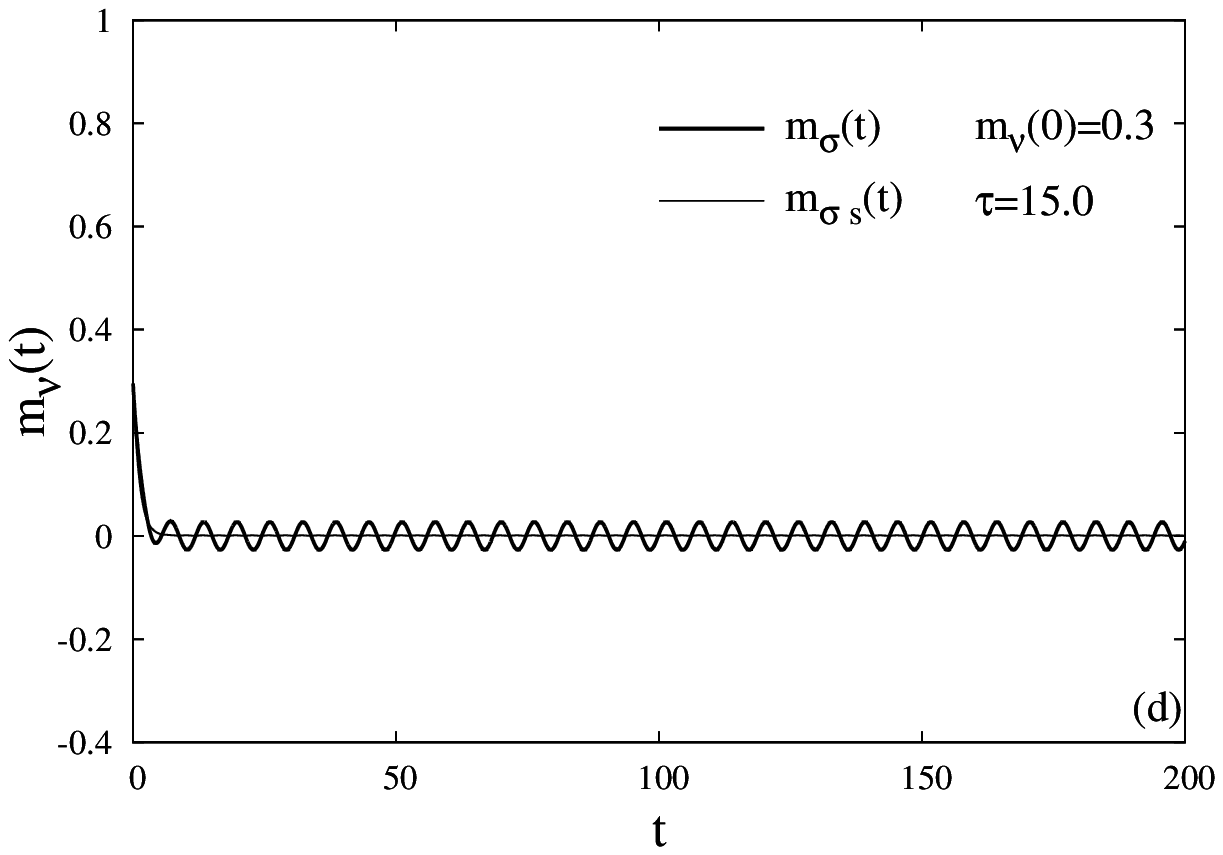, width=7.2cm}
\end{center}
\caption{Time series of the order parameters of the IATM which corresponding to 
(a) DP1, (b) DP4, (c) DP3 and (d) DP2. All time series are calculated for 
the Hamiltonian parameters $K_4=0.5, \omega=1.0, h_0=0.5$. Initial values of the 
order parameters and temperature are shown in each figure. In each figure, 
the thicker lines correspond to $m_\sigma=m_s$ and the thinner lines correspond to 
$m_{\sigma s}$. $\nu$ stands for  $\sigma, s, \sigma s $. } 
\label{sek1}\end{figure}

These phases and borders between them can be determined by calculating 
magnetizations defined in Eq. \re{denk10} by using formulation presented in this 
work. Typical time series corresponding to these four different phases can be seen 
in  Fig. \ref{sek1}. 

Although the values of the magnetizations in the DP1, DP2 and DP3 phases are not 
dependent on the initial values of the magnetizations, as one can see from to Fig. 
\ref{sek1} (b) that, this does not hold for the DP4 phase. The value of the 
$m_\sigma=m_s$ converges zero or specific nonzero values depending on the 
initial value. In other words, formulation gives two different values (such that 
one of them is zero) for $m_\sigma=m_s$ depending on the choice of the 
initial values.

\begin{figure}[h]\begin{center}
\epsfig{file=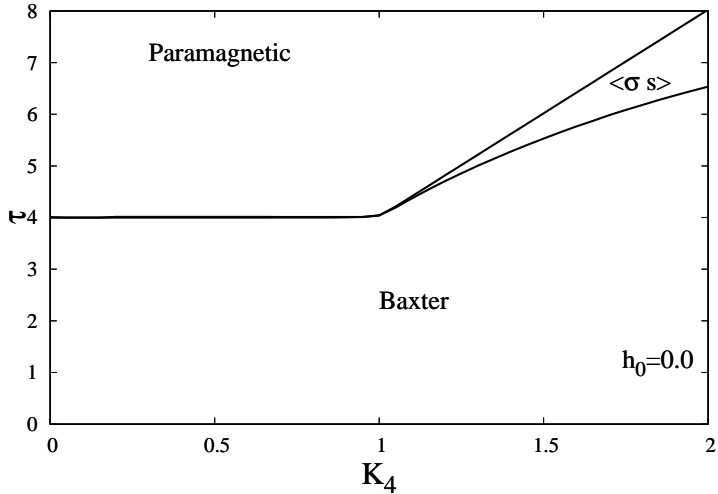, width=10.0cm}
\end{center}
\caption{Phase diagram of the static IATM in the $(K_4,\tau)$ plane. } 
\label{sek2}\end{figure}

The phase diagram for static IATM in $(K_4,\tau)$ plane can be seen in Fig. 
\re{sek2}. This diagram can be obtained by numerical solutions of the Eq. 
\re{denk11} in the static limit. The same phase diagram obtained within the RG 
\cite{ref19,ref20} and MC \cite{ref25} can be found in the literature.  In the static IATM, 
for the higher values of the  $K_4$, $\sandd{\sigma s}$ phase may appear when 
the temperature rises, as seen in Fig. \ref{sek2}.

\subsection{Dynamic Phase Boundaries}

Dynamic Phase Boundaries (DPB) of the model separate the different dynamic 
phases of the model. The relation of the dynamic behavior of the system with 
driving dynamical magnetic field is well known \cite{ref59}. The relation 
between the exchange interaction, temperature, amplitude and the frequency of 
the magnetic field determines the dynamic phase of the system. Ferromagnetic 
exchange interaction of the model tends to to keep the nearest neighbors of the spins 
of the system parallel to each other.  Rising temperature gives rise to the thermal thermal 
fluctuations and drive the system to the disordered phases. Besides, in the dynamical 
case, the amplitude and the frequency of the magnetic field are also decisive. 
Higher amplitude, dictates the spins to align parallel to the field. But when the 
frequency is high, spins cannot follow to magnetic field due to the difference 
between the period of the magnetic field and relaxation time of the spins.

\begin{figure}[h]\begin{center}
\epsfig{file=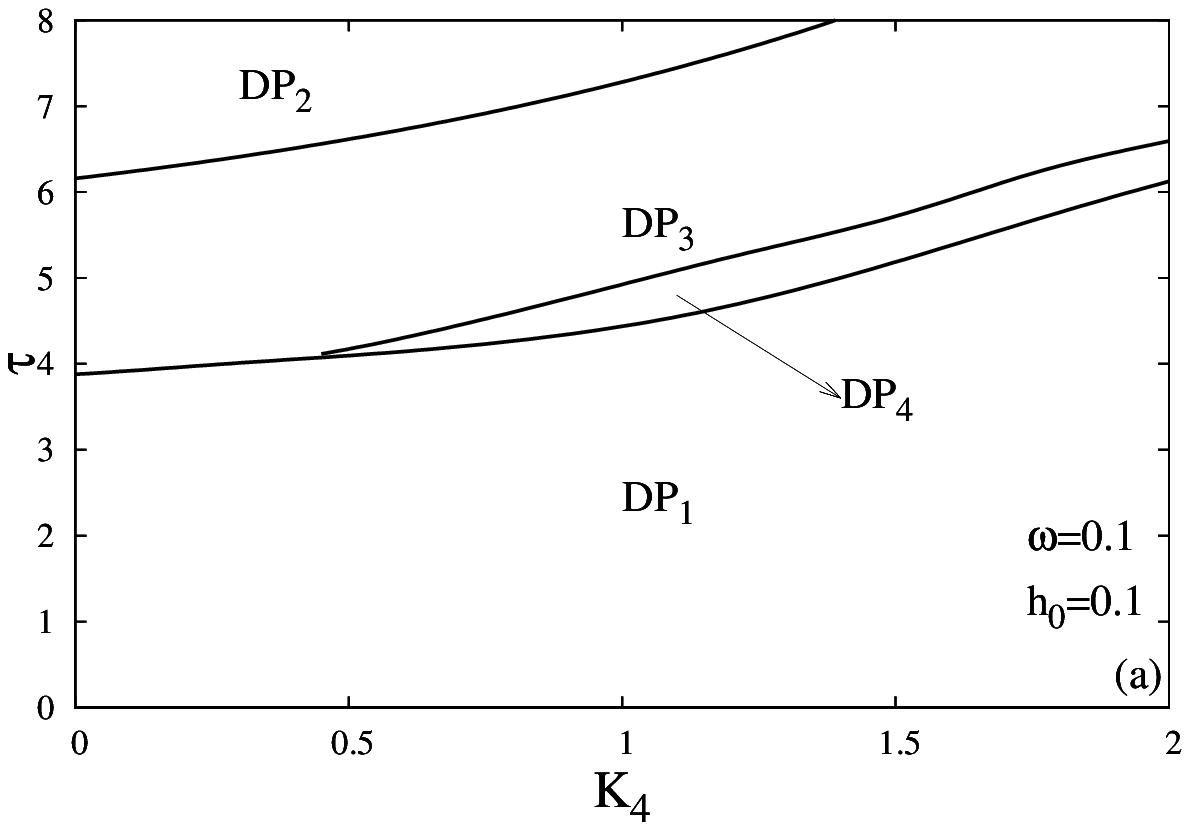, width=7.2cm}
\epsfig{file=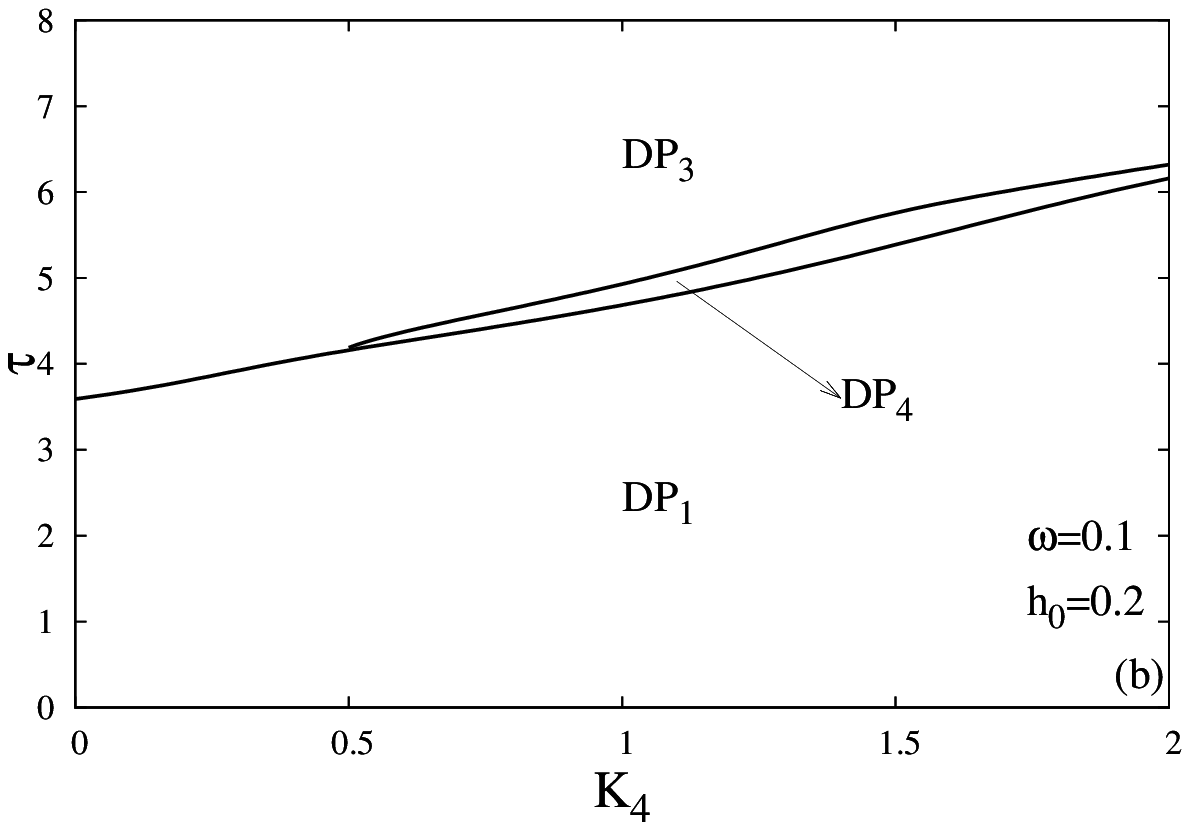, width=7.2cm}
\epsfig{file=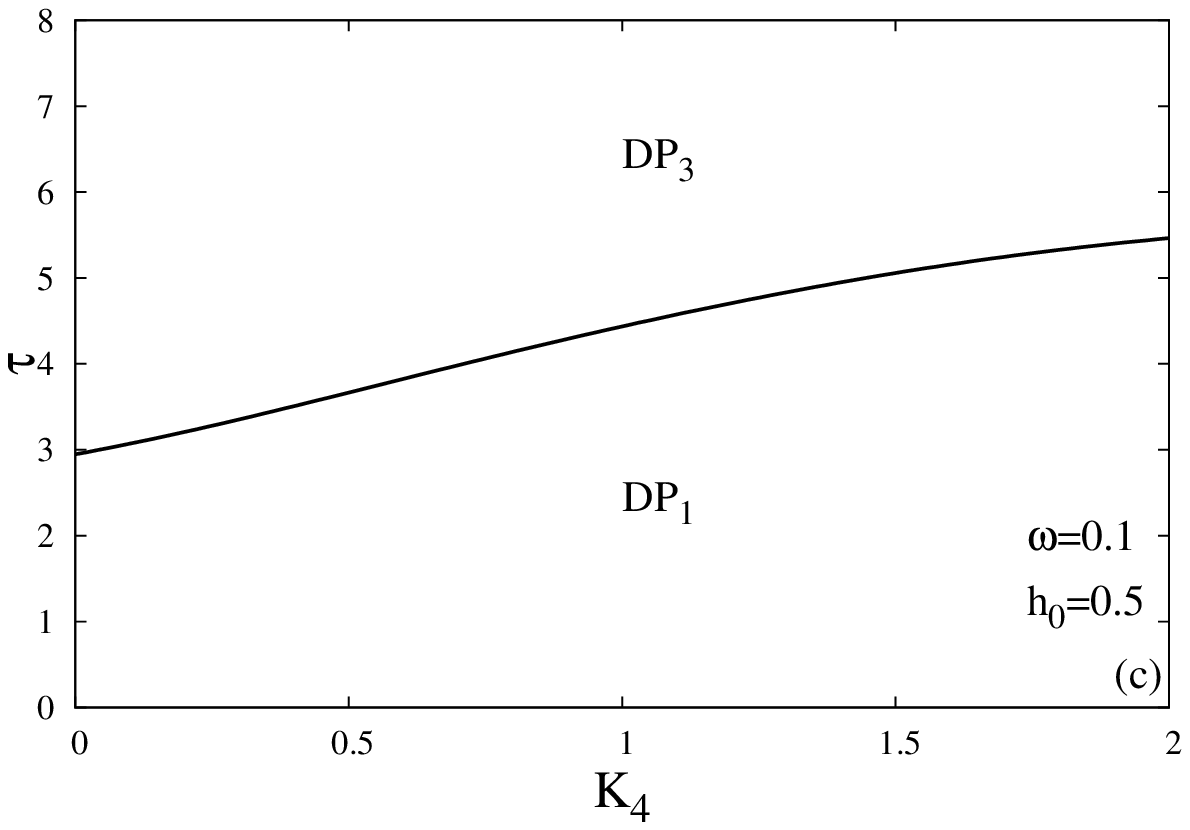, width=7.2cm}
\epsfig{file=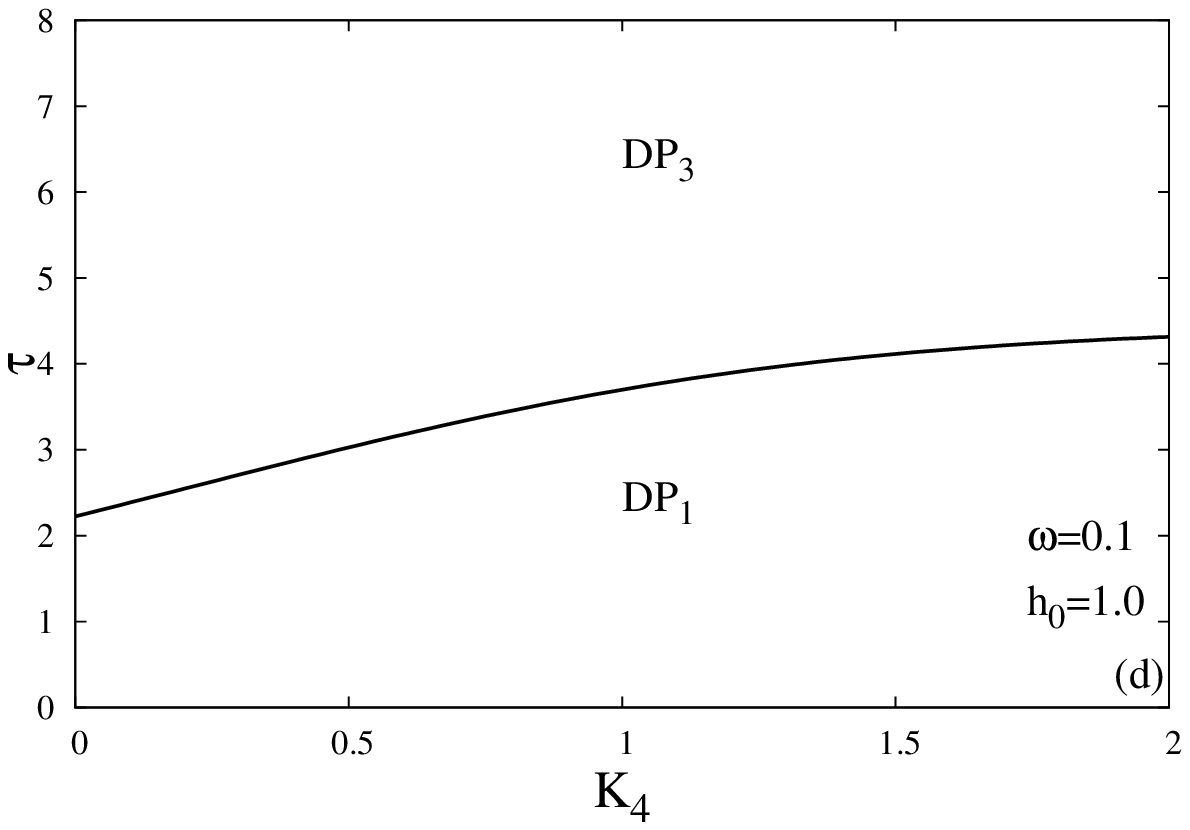, width=7.2cm}
\end{center}
\caption{Phase diagrams of the dynamical IATM in a $(K_4,\tau)$ plane for selected 
values of $\omega=0.1$ and (a) $h_0=0.1$, (b) $h_0=0.2$, (c) $h_0=0.5$, (d) 
$h_0=1.0$. } \label{sek3}\end{figure}

\begin{figure}[h]\begin{center}
\epsfig{file=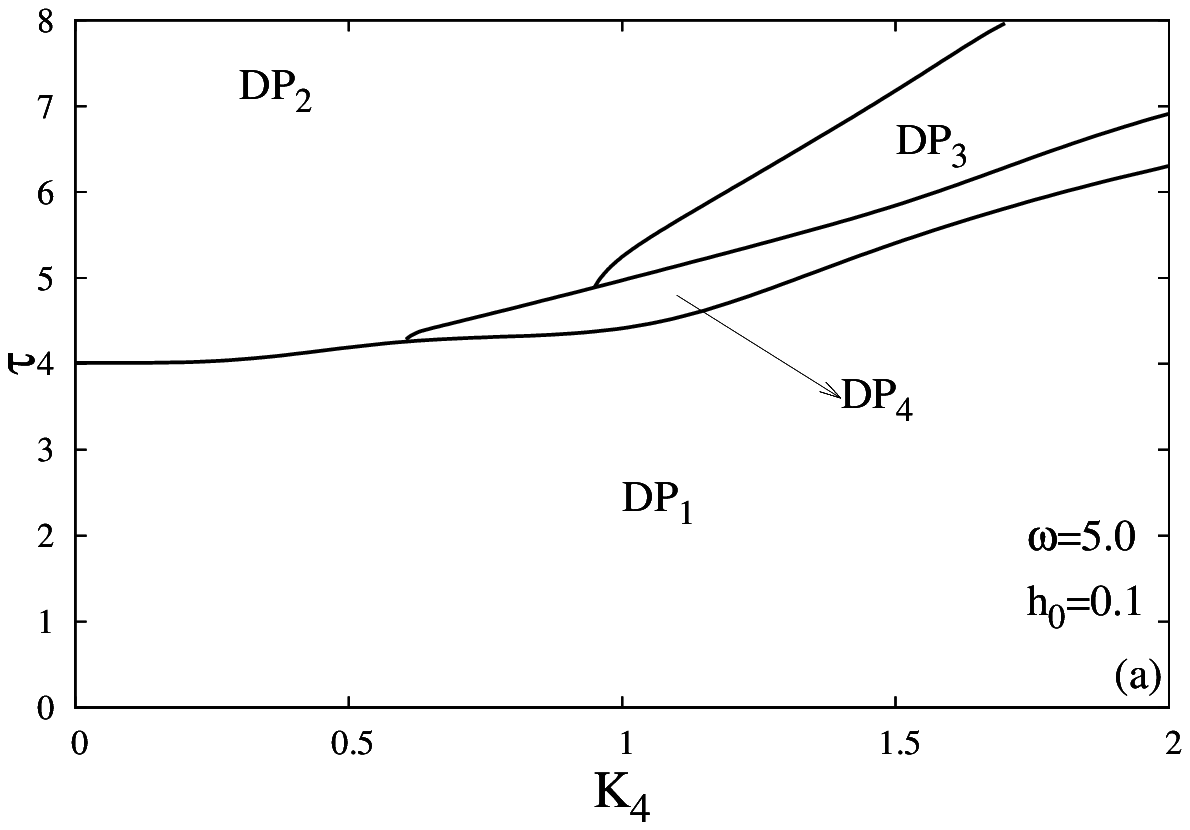, width=7.2cm}
\epsfig{file=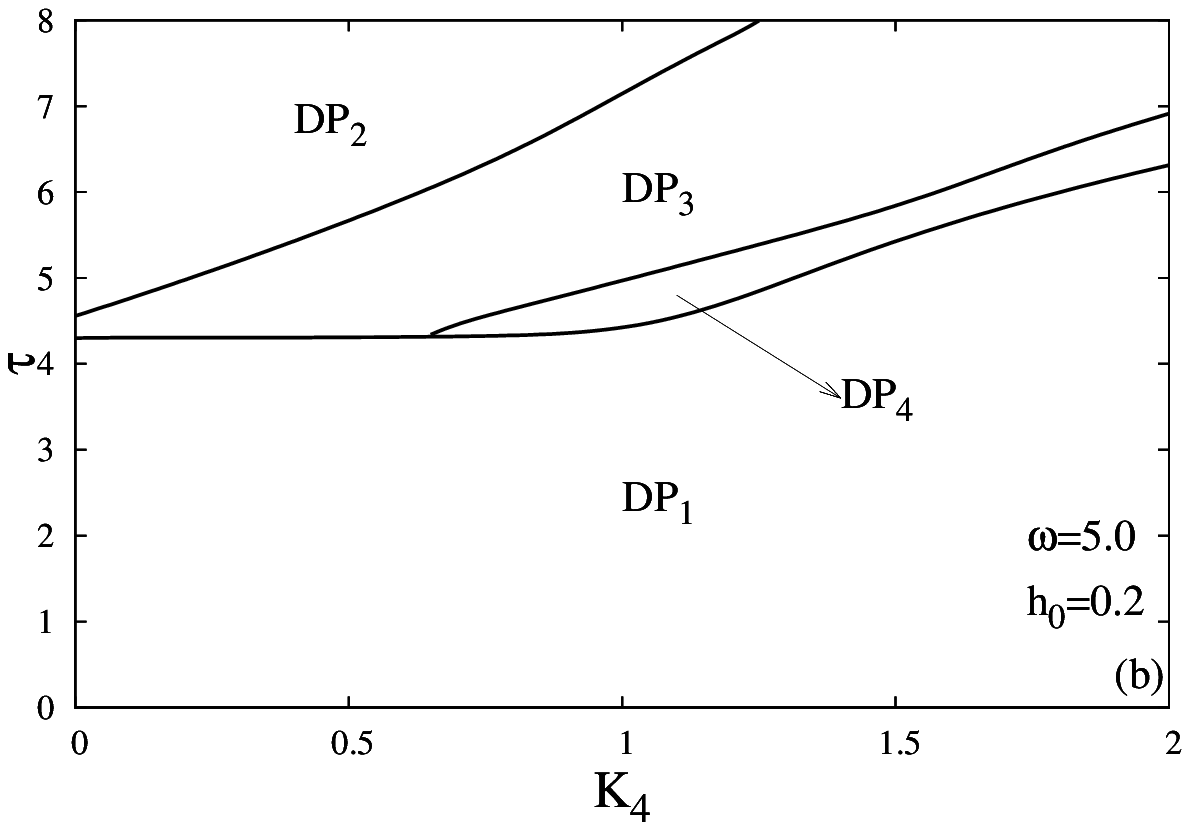, width=7.2cm}
\epsfig{file=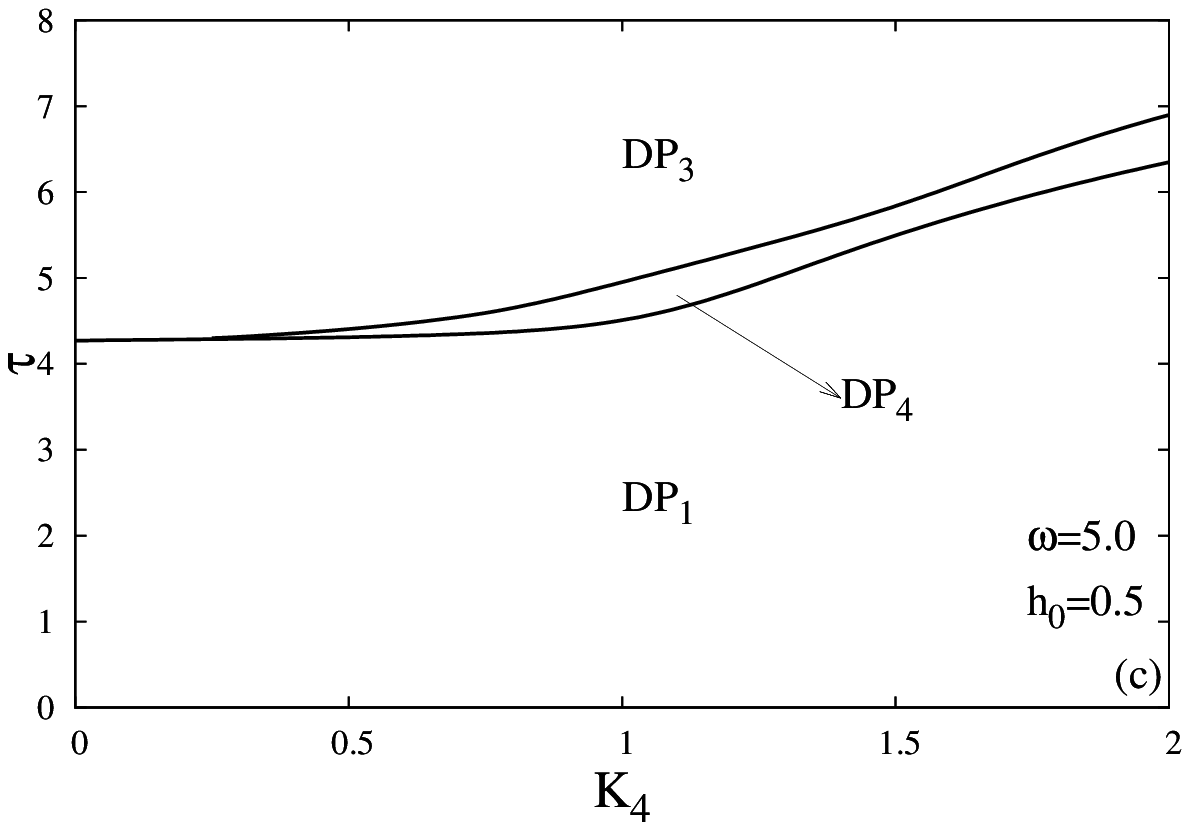, width=7.2cm}
\epsfig{file=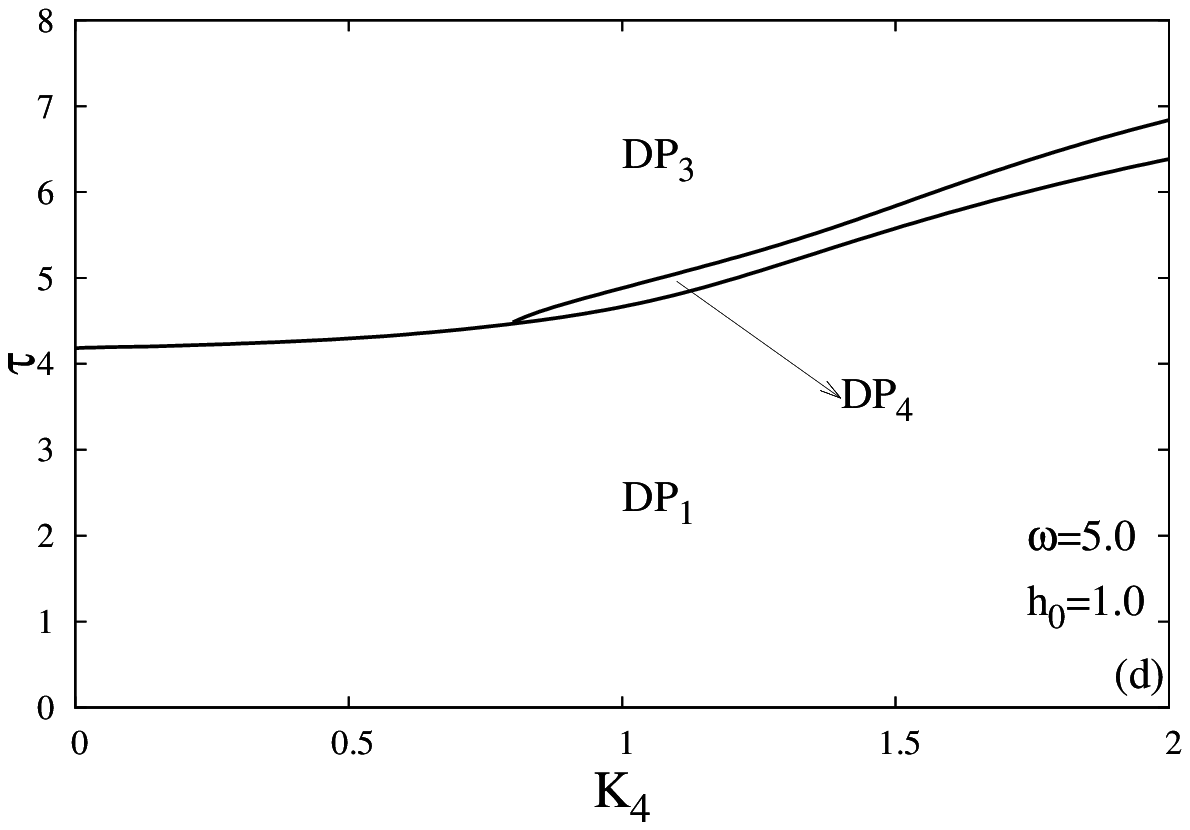, width=7.2cm}
\end{center}
\caption{Phase diagrams of the dynamical IATM in a $(K_4,\tau)$ plane for selected 
values of $\omega=5.0$ and (a) $h_0=0.1$, (b) $h_0=0.2$, (c) $h_0=0.5$, (d) 
$h_0=1.0$. } 
\label{sek4}\end{figure}

In order to compare with the static case, let us depict the phase diagrams in 
$(K_4,\tau)$ plane (as in Fig. \ref{sek2}) for some selected values of $\omega$ 
and $h_0$. This can be seen in Fig. \ref{sek3} for low frequency ($\omega=0.1$) 
and in Fig. \ref{sek4} for high frequency ($\omega=5.0$), with selected values of 
$h_0=0.1,0.2,0.5,1.0$ shown in figures labeled as (a),(b),(c) and (d) 
respectively.

As seen in Fig. \ref{sek3}, at low frequencies, rising amplitude shrinks the 
$DP_1$ region and shifts the $DP_2$ region towards the higher temperature 
region of the $(K_4,\tau)$ plane. Also disappearing of the $DP_4$ phase with 
rising  $h_0$ draws attention. The same reasoning holds for the high frequency 
regime, which can be seen in Fig. \ref{sek4}. For higher frequency the $DP_4$ 
phase survives in a larger interval of $h_0$, e.g. while for the value of 
$h_0=1.0$, high frequency has $DP_4$ phase in $(K_4,\tau)$ plane (Fig. 
\ref{sek4} (d)), the same phase don't occur for lower frequency (Fig. \ref{sek3} 
(d)). This behavior is similar to the behavior of the dynamical Ising model as 
shown in  \cite{ref65}. 

\subsection{Hysteresis characteristics}

The hysteric response of the system to the varying temperature, frequency and 
amplitude of the driven field is well known from the investigations on the Ising 
model \cite{ref59}. Thus, in this section we especially want to discuss the 
effect of the coupling constant $K_4$ on these hysteresis characteristics.

\begin{figure}[h]\begin{center}
\epsfig{file=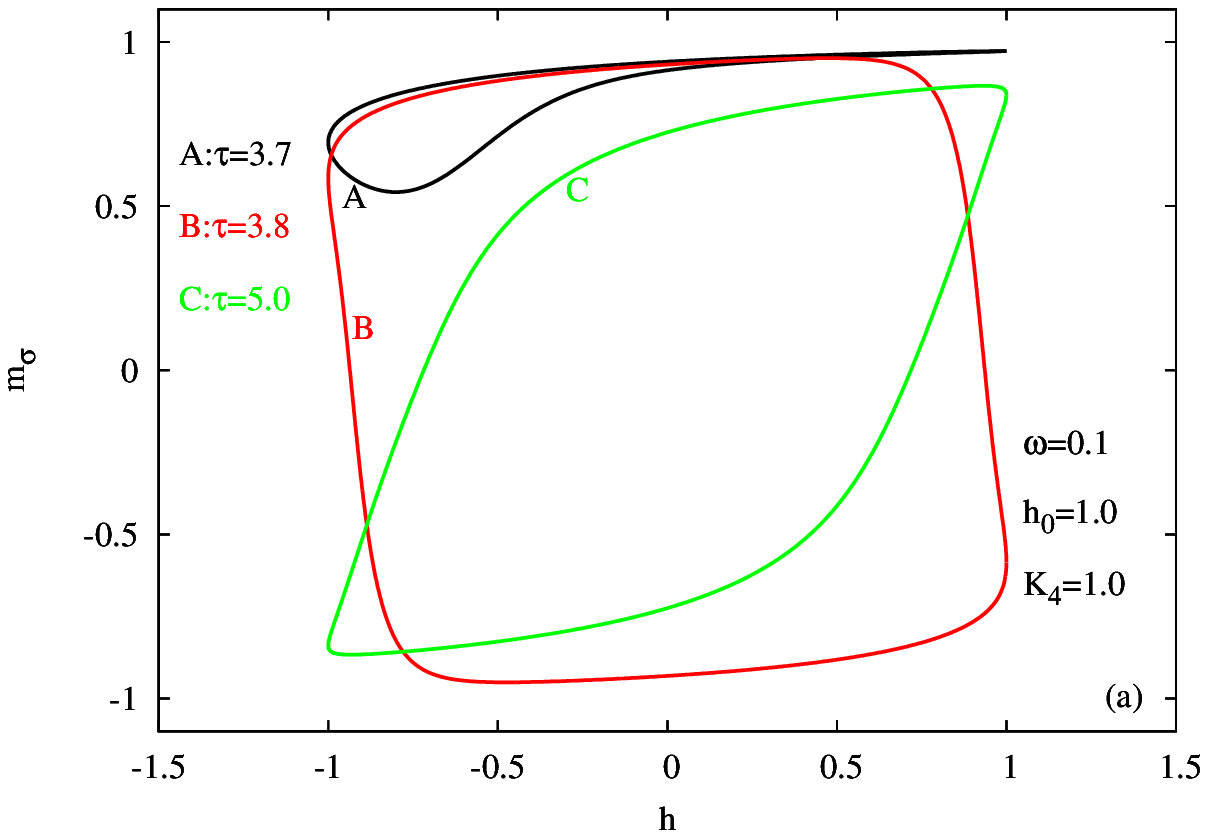, width=7.4cm}
\epsfig{file=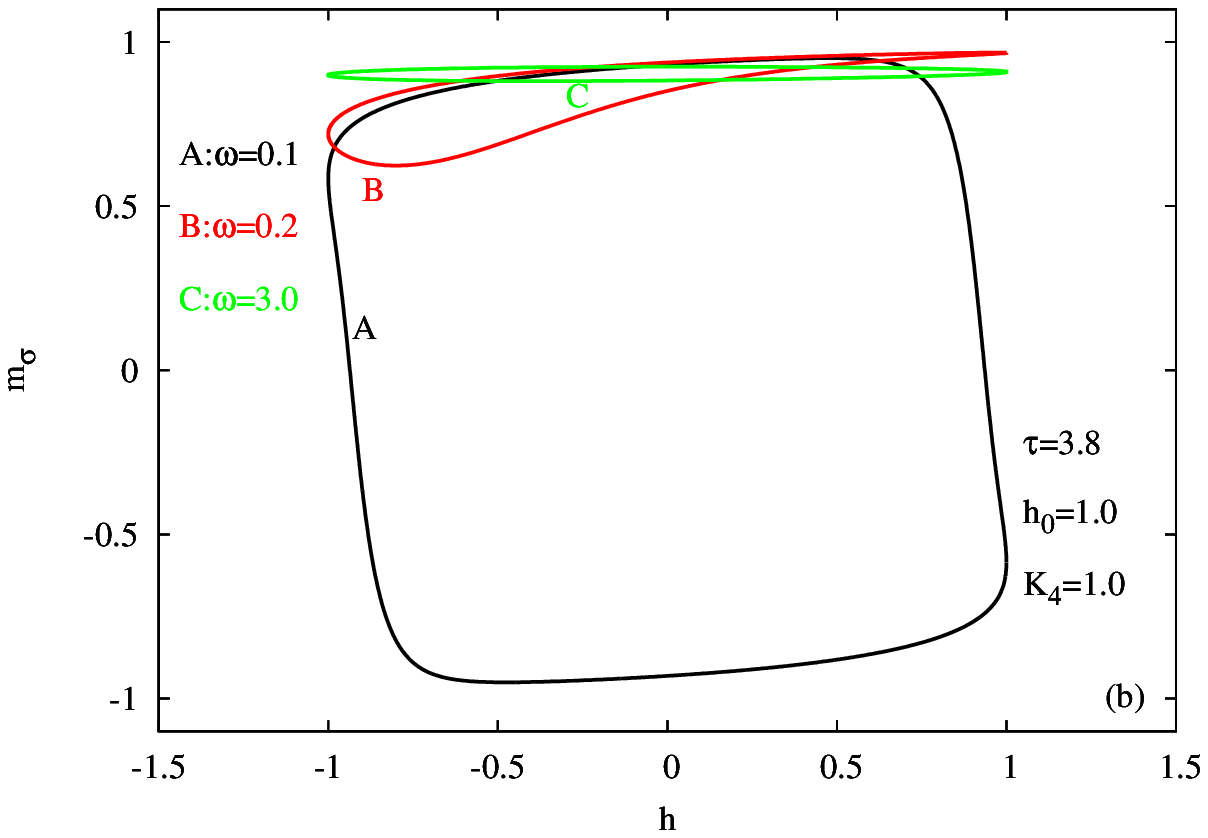, width=7.4cm}
\epsfig{file=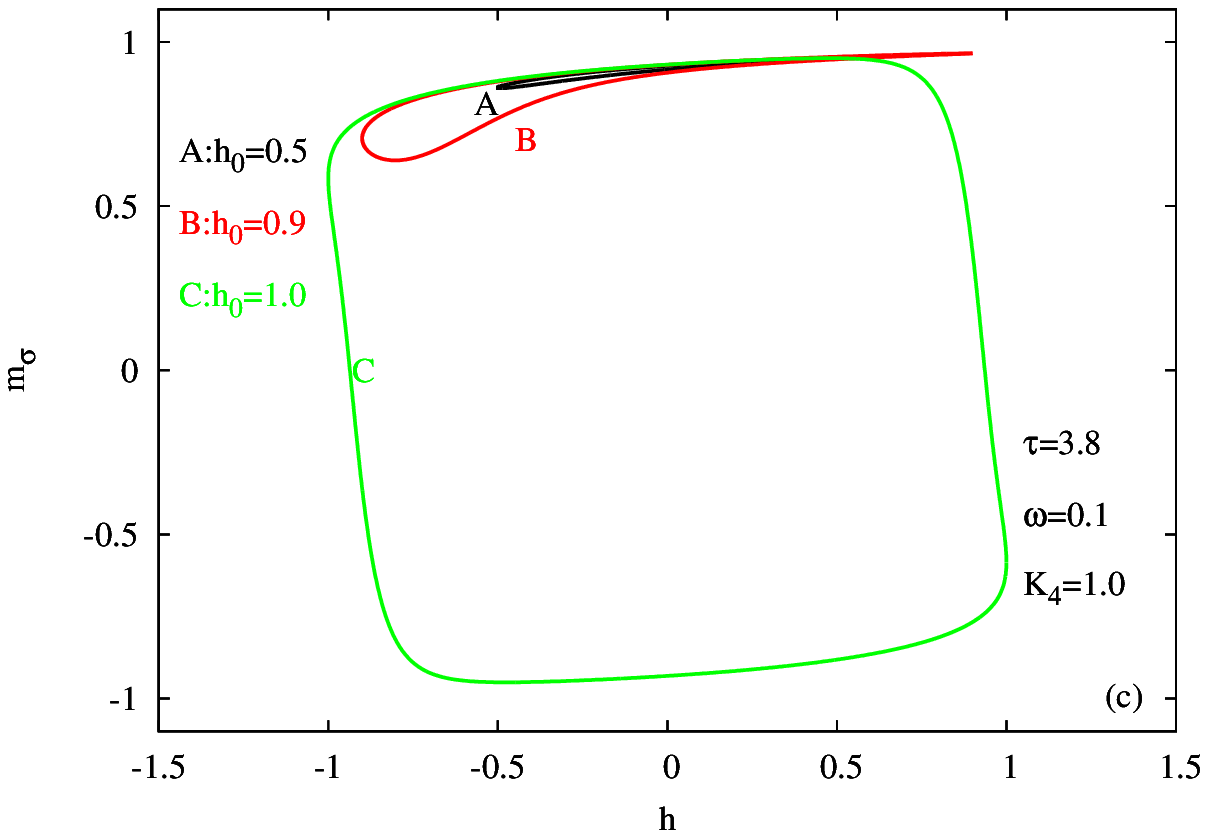, width=7.4cm}
\epsfig{file=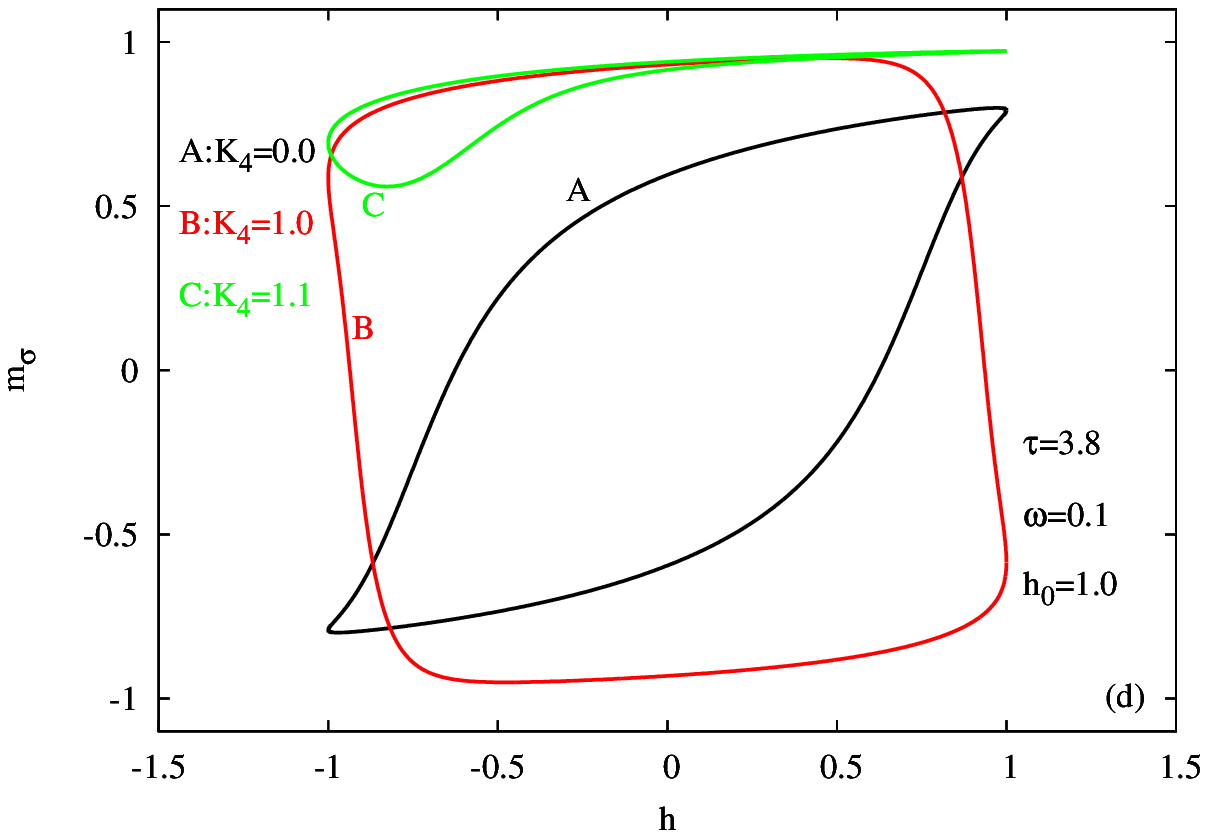, width=7.4cm}
\end{center}
\caption{Hysteresis behavior of the IATM on a square lattice for changing (a)  
temperature, (b) frequency of the field, (c) amplitude of the field and (d) $K_4$, 
for selected values of Hamiltonian parameters.} \label{sek5}\end{figure}

Typical behaviors of the hysteresis loops with changing Hamiltonian parameters 
can be seen in Fig. \ref{sek5}. If we group hysteresis loops (as for the Ising 
model) as paramagnetic and ferromagnetic loops, we observe from Fig. \ref{sek5} that, in the manner of 
transition between two kind of loops  

\begin{itemize}

\item The effect of the rising temperature and amplitude of the field are 
similar, they give rise to transition from ferromagnetic loops to 
paramagnetic loops (see Fig. \ref{sek5} (a) and (c))
\item The effect of the rising  frequency of the field and $K_4$ are similar, 
they give rise to transition from paramagnetic loops to ferromagnetic loops (see Fig. \ref{sek5} (b) and (d)).

\end{itemize} 

Rising temperature causes enhanced thermal fluctuations and the spins can 
follow the driving periodic magnetic field. The same reasoning holds for the rising 
amplitude of the field, since rising amplitude of the field means that more 
energy is supplied to the system from the magnetic field. These two mechanisms 
cause a transition from the ordered phase to the disordered phase. The reverse 
transition can be seen for the rising frequency of the field. After a specific  
frequency (which depends on all other Hamiltonian parameters), spins cannot 
follow the magnetic field, then ferromagnetic phase occurs. These facts have already 
been observed and explained for the dynamical  Ising model \cite{ref59}. For the 
IATM we observe from Fig. \ref{sek5} (d) that rising $K_4$ can give rise to 
transition from the paramagnetic hysteresis loops to the ferromagnetic loops 
(compare loops labeled by A,B,C in Fig. \ref{sek5} (d)). This behavior is 
consistent with the dynamic phase diagram which was depicted in Fig. \ref{sek3} 
(d).  As seen in Fig. \ref{sek3} (d) while system is in the $DP_3$ phase for the 
temperature value of $\tau=3.8$, rising coupling constant $K_4$ gives rise to a 
transition to the $DP_1$ phase which has ferromagnetic hysteresis 
characteristics.

\begin{figure}[h]\begin{center}
\epsfig{file=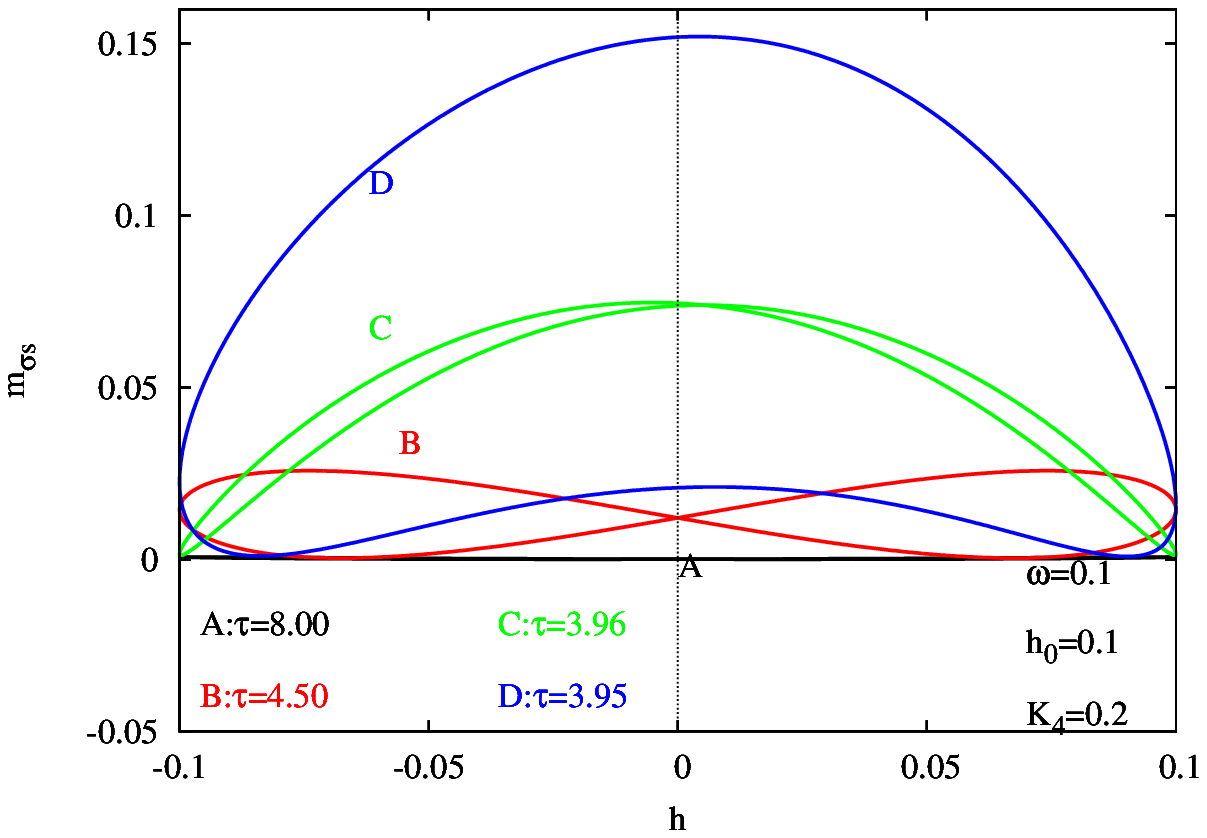, width=12cm}
\end{center}
\caption{Variation of $m_{\sigma s} $ within the one period of the magnetic 
field for selected temperature values as $\tau=8.00,4.50,3.69,3.65$. Other 
values of the Hamiltonian parameters are fixed as $\omega=0.1,h_0=0.1, K_4=0.2$.  } 
\label{sek6}\end{figure}

Note that there is not any significant difference between the observed hysteresis loops belonging to $DP_2$
 and $DP_3$ phases. This is because both of the 
phases have $m_\sigma=0$. Thus the phases $DP_2$ and $DP_3$ could not be 
distinguished by  hysteresis behaviors of order parameter $m_{\sigma}$. But the variation of the $m_{\sigma s} $ 
with magnetic field in one period may distinguish between these two phases. This loop may be 
called as hysteresis loop of $m_{\sigma s} $. In order to determine the response 
of this new order parameter (which is absent in the Ising model), we depict 
typical loops representing the phases $DP_2, DP_3$ and $DP_1$ in Fig. \ref{sek6}. 
The other Hamiltonian parameter values are chosen as $\omega=0.1, 
h_0=0.1,K_4=0.2$.  We can see from Fig. \ref{sek3} (a) that, the curve 
labeled A in Fig. \ref{sek6}  represents the hysteresis loop of the phase 
$DP_2$, B represents the phase $DP_3$ and C and D correspond to the phase 
$DP_1$. First of all, the loop corresponds to the phase $DP_2$ (curve labeled by 
A) indeed a $m_{\sigma s}=0$ line, means that the order parameter $m_{\sigma s}=0$ does not 
respond to the varying magnetic field. As seen in Fig. \ref{sek6}, the order 
parameter $m_{\sigma s}$ starts to reply the changing magnetic field within 
the phase $DP_3$ (curve labeled B) and similar loop is valid for the phase 
$DP_1$ (curve labeled C). But one difference draws our attention: while the 
loop belonging to the phase $DP_3$ has one $m_{\sigma s}$ value for the zero 
field, the loop about the $DP_1$ phase has two very close but slightly different 
values of $m_{\sigma s}$ for the $h=0$ case. When the temperature is lowered, the loop 
like knot becomes dissolved and evolve into the loops like labeled D in Fig. \ref{sek6}. As 
a result, we can distinguish the phases $DP_1,DP_2$ and $DP_3$ by looking at the 
hysteresis loops of the $m_{\sigma s}$ as explained above.

\begin{figure}[h]\begin{center}
\epsfig{file=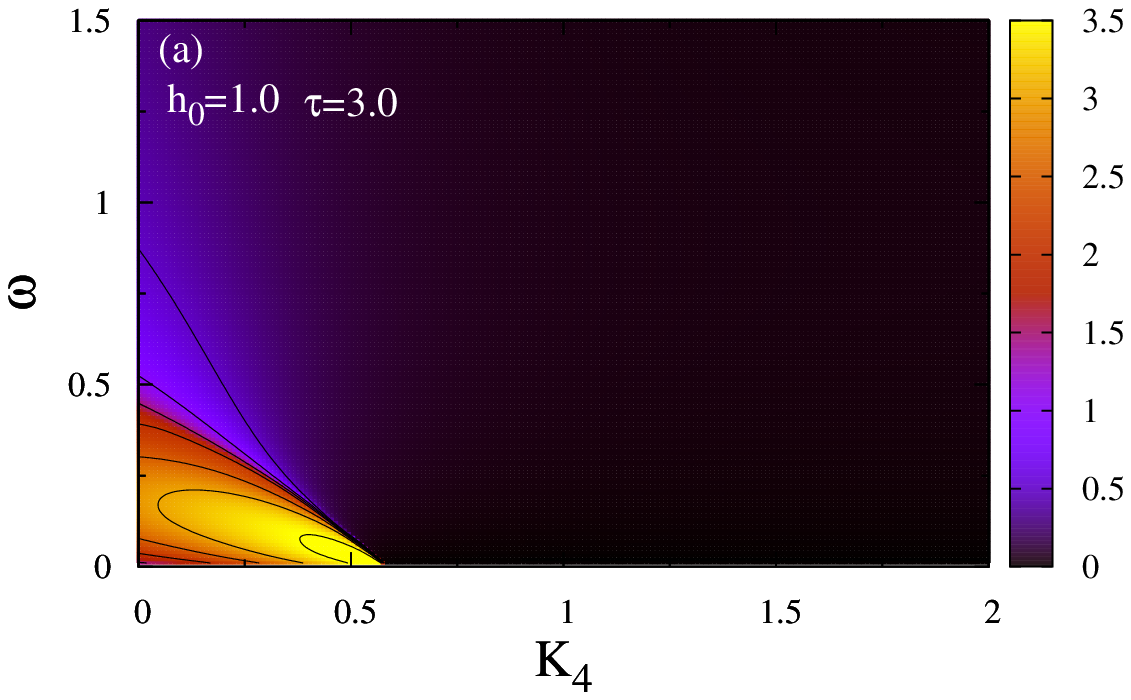, width=7.4cm}
\epsfig{file=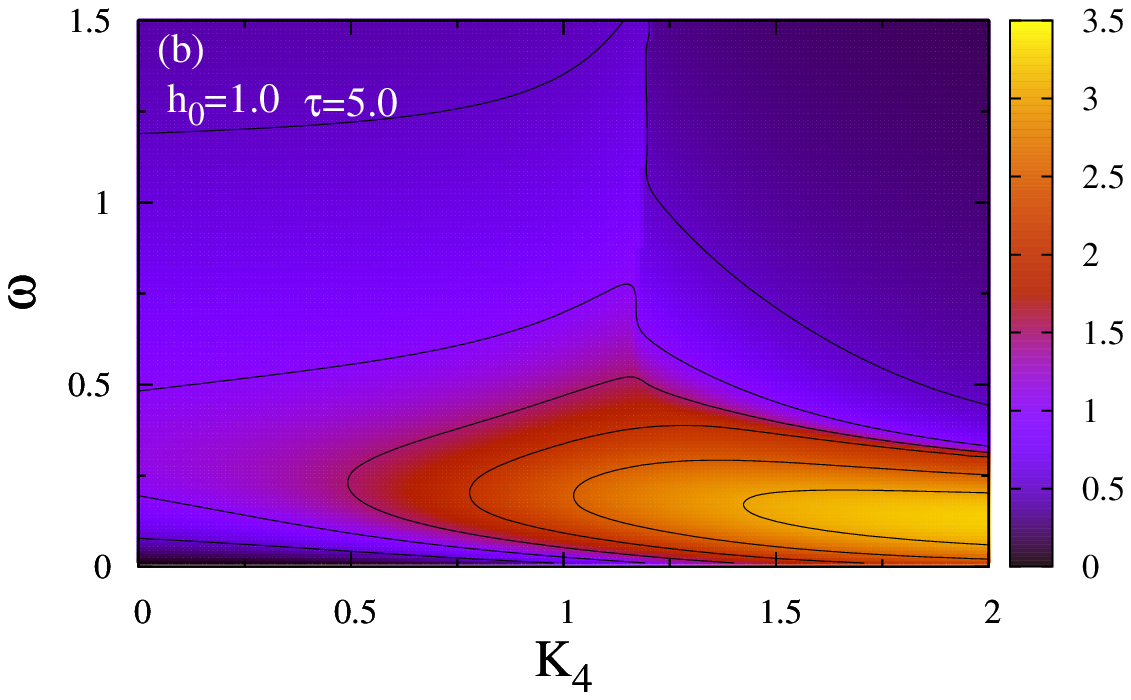, width=7.4cm}
\epsfig{file=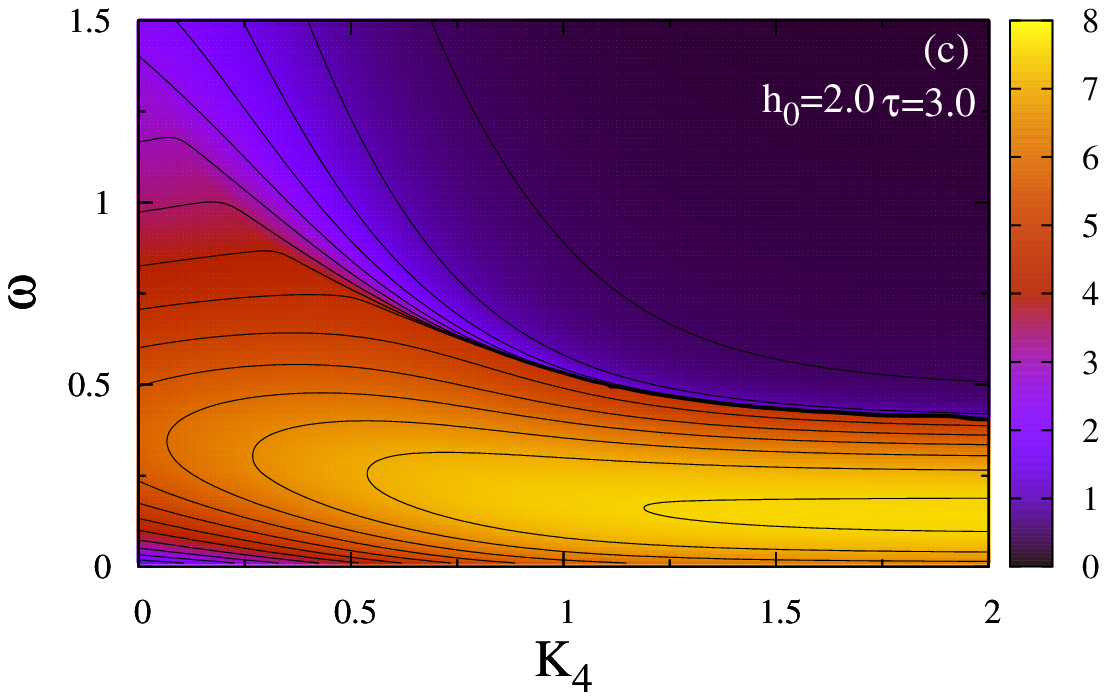, width=7.4cm}
\epsfig{file=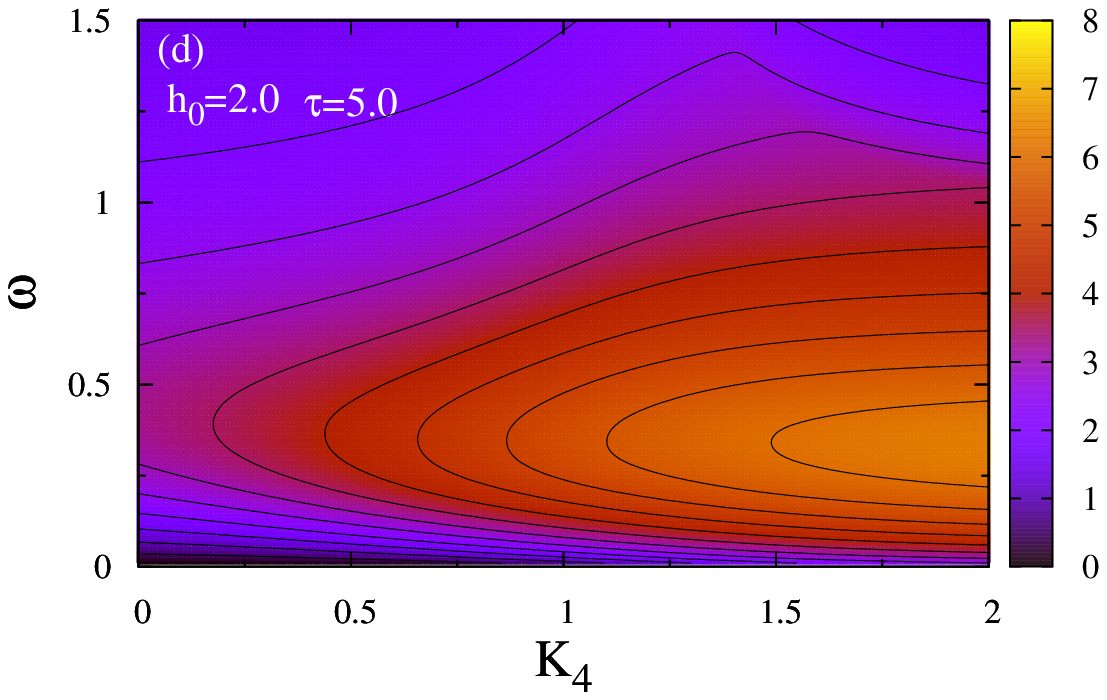, width=7.4cm}
\end{center}
\caption{Contour plots of the HLA in ($\omega,K_4$) plane for selected values of 
$(h_0,\tau)$ pair as (a) ($1.0,3.0$) (b) ($1.0,5.0$) 
(c) ($2.0,3.0$) (d) ($2.0,5.0$). Also contour lines added as starting with the 
value of $0.5$ and increment of $0.5$. }  \label{sek7}\end{figure}

Lastly we want to elaborate the HLA properties of the system. For this aim we depict the 
contour plots of the HLA in ($\omega,K_4$) plane for selected values of 
$(h_0,\tau)$ pair, which can be seen in Fig. \ref{sek7}. HLA of the dynamical Ising 
model has been widely inspected in the literature and the variation of the HLA with the 
frequency is well known. We want to determine the relation between  $K_4$ and 
the HLA. This can be seen   in contour plots for several cases. At first sight 
one difference is take our attention between the contour plots given in  Fig. 
\ref{sek7} (a)-(d). Rising $K_4$ with low frequencies result in lowering of the 
HLA in Fig. \ref{sek7} (a) while the reverse relation holds in Figs. \ref{sek7} 
(b)-(d). As seen in Fig. \ref{sek3} (d), rising $K_4$ give rise to transition 
from the phase $DP_3$ to $DP_1$ and this transition causes depressed HLA. We 
cannot face this situation in Fig. \ref{sek7} (b). For low frequencies, rising 
$K_4$ enhances the HLA. Again from Fig. \ref{sek3} (d), we can see that, this 
variation cannot change the phase of the system, the system lies in the $DP_3$ 
phase and rising coupling $K_4$ between the two Ising models give rise to 
rising HLA. Same situation holds for Figs. \ref{sek7} (c) and (d). The only 
difference is wider region of the $(K_4,\omega)$ plane has big HLA due to the 
rising $h_0$ enlarges the region that have $DP_2,DP_3$ regions.        

\section{Conclusion}\label{conclusion}

The phase diagrams and the hysteresis characteristics of the IATM has been 
investigated within the MFA with Glauber type of stochastic process. 
First, the phase diagrams of the model have been obtained in the $(\tau,K_4)$ plane, by 
defining the possible dynamical phases of the system obtainable within the used 
approximation. The effect of the frequency and the amplitude of the periodic 
time dependent magnetic field on these diagrams has been investigated in 
detail. 

To the hysteresis part of the work, since the effect of the Hamiltonian 
parameters on the hysteresis characteristics on the Isng model is well known and
the effect of the Hamiltonian parameter $K_4$ on the hysteresis characteristics 
is mostly investigated. Dynamical phase transitions induced by changing $K_4$ 
shows itself on the hysteresis behaviors. These behaviors have been discussed in detail. 
Besides, the behavior of the  order parameter of the IATM  namely $m_{\sigma s}$ 
with the magnetic field has also been discussed. The relation between these hysteresis loops 
and possible phase transitions explored.

We hope that the results  obtained in this work may be beneficial 
form both theoretical and experimental point of view.

\end{document}